\documentclass[preprint2]{aastex}
\slugcomment{to be published in Astrophysical Journal Letters}
\lefthead{Liang et al.}
\righthead{IR Line Emission in the Antennae}
\begin{document}
\title{Infrared Line Emission in the Interacting Region of Arp 244 (the
 Antennae): Colliding Molecular Cloud Complexes ?}
\author{M. C. Liang\altaffilmark{1}}
\affil{Institute of Astronomy and Astrophysics, Academia Sinica, P. O.
 Box 1-87, Nankang, Taipei 115, Taiwan}
\email{danie@asiaa.sinica.edu.tw}
\and
\author{T. R. Geballe}
\affil{Gemini Observatory, Hilo HI  96720}
\email{tgeballe@gemini.edu}
\and
\author{K. Y. Lo, \& D.-C. Kim}
\affil{Institute of Astronomy and Astrophysics, Academia Sinica, 
 P. O. Box 1-87, Nankang, Taipei 115, Taiwan}
\email{kyl@asiaa.sinica.edu.tw, kim@asiaa.sinica.edu.tw}
\altaffiltext{1}{Dept. of Physics, Univ. of Tsing Hua, 101, Section 2 Kuang 
 Fu Road, Hsinchu 300, Taiwan}

\begin{abstract}

We report velocity-resolved spectroscopy of infrared hydrogen
recombination lines in the interacting region of the Antennae galaxies
(NGC 4038/4039). It generally has been assumed that the active star
formation found there is due to the interaction of the disks of the two
galaxies and indeed two molecular cloud complexes,
separated in velocity by $\sim$100 km s$^{-1}$, have been observed 
in the southern part of
this region. Our measurements imply that the two cloud complexes are
moving away from each other.  This result poses interesting questions
about the physical mechanisms leading to starbursts in Arp 244 and other
interacting galaxies.

\end{abstract}


\section{INTRODUCTION}

The Antennae (NGC 4038/4039, Arp 244, VV245) are a nearby ($\sim$22 Mpc)  
infrared-luminous pair of interacting galaxies. Using N-body numerical
simulations \citet{tt72} succeeded in generating the long tidal tails
seen in optical/\ion{H}{1} images of the galaxies \citep{hulst79}. The
strongest emission in both the radio continuum \citep{Hvan86} and the
CO(1-0) line observed by \citet{SM85, stanfordetal90, aaltoetal95} and by
S. W. Lee,
K. Y. Lo, R. A Gruendel, and Y. Gao (2001, in preparation, hereafter L01)
occurs in
the interaction (overlap) region to the west and between the two nuclei,
especially in its southern dense clouds. The Hubble Space Telescope
H$\alpha$ image of the Antennae reveals thousands of giant \ion{H}{2}
regions in the interaction region. X-ray observations \citep{FT83} show
harder and softer components corresponding to SNRs/X-ray binaries and
thermal emission from gas possibly heated by the interaction,
respectively. These observations all point to there being copious numbers
of active star formation sites in the interacting region.

Mid-infrared ($12-17~\micron$) observations \citep{m98} of the Antennae by
the Infrared Space Observatory (ISO) confirm that the starburst is taking
place largely in portions of the interaction region which are highly
optically obscured.  The ratio of [\ion{Ne}{3}] and [\ion{Ne}{2}]
mid-infrared lines indicates that stars as massive as 60 M$_{\sun}$ are
present there \citep{vig96,m98}. The peak of [\ion{C}{2}] 158 $\micron$
fine-structure line also lies in the interaction region, but overall the
ratio of [\ion{C}{2}] to CO is a factor of 2.6 lower than usual starburst
galaxies, suggesting that global star formation is not taking place in the
Antennae \citep{nikola98}.

\begin{figure*}
\epsscale{1.19}
\plotone{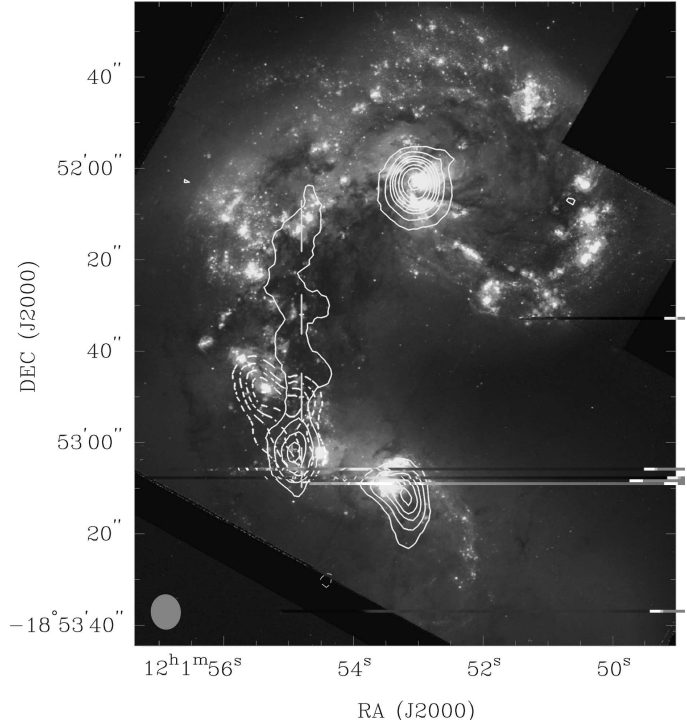}
\caption[CO(1-0) contour on HST 3-color image]{The CO(1-0) moment-zero low
resolution (7.8$\arcsec \times$6.6$\arcsec$) contour map (L01)
overlayed on the HST 3-color (H$\alpha$, V-band, and B-band) image. The
long vertical crosses indicate the locations along the slit where IR
recombination lines were detected; from top to bottom these are
Apertures I-IV.  Solid contours represent the red
component of the CO emission in the interaction region and the CO
emissions in each nuclei (contour levels are 6 Jy beam$^{-1}$ km s$^{-1}$
times 4, 7, 10, 12, 14, 17, 20, and 25). Dashed contours are the blue
component of the CO emission in the interaction zone (levels are 5 Jy
beam$^{-1}$ km s$^{-1}$ times 5, 9, 13, 17, and 21).
\label{fig-1}}
\end{figure*}

Using the BIMA Array with a 7.8$\arcsec$~$\times$~6.6$\arcsec$ beam
L01 have found that there are two kinematically distinct
molecular cloud complexes separated by $\sim$100 km s$^{-1}$ in the
southern part of the overlap region (see Fig.~\ref{fig-1}), corresponding
roughly to super giant molecular complexes 4 and 5 found by
\citet{wilson00}. It seems likely that each of these is associated with
one of the interacting galaxies.  Because their relative positions along
the line of sight are unknown, it is unclear if the complexes are
approaching or moving away from each other. If the redshifted cloud
complex is in front, it may collide with blueshifted cloud and
subsequently a starburst, perhaps greater than that currently being
observed, may occur. If the blueshifted clouds are in front, then the
complexes are moving away from each other and the bulk of their
interaction, if any, already has taken place in this region.

Comparison of velocity profiles of properly chosen emission lines, well
separated in wavelength but from similarly excited species, could clarify
the situation, as velocity components associated with the two complexes
should undergo different amounts of extinction. UV and optical lines
associated with star formation in these clouds are expected to be
essentially totally obscured by dust. With this in mind we have observed
the hydrogen infrared recombination lines Pa$\beta$ (5-3, 1.282~$\mu$m),
Br$\gamma$ (7-4, 2.166~$\mu$m), and Br$\alpha$ (5-4, 4.052~$\mu$m) from
part of the overlap region.

\section{OBSERVATIONS AND DATA REDUCTION}

The observations were made during UT 2000 28 February--1 March at the
United Kingdom Infrared Telescope (UKIRT) on Mauna Kea, using the facility
cooled grating spectrometer CGS4 \citep{m90} equipped with a
256$\times$256 InSb array.  All three nights were clear with stable, but
not exceptional seeing. We selected the 31 l/mm echelle, 300~mm focal
length camera optics, and 2-pixel wide (0.8$\arcsec$) slit, which give
CGS4 a velocity range of $\sim$2,000~km~s$^{-1}$ at any wavelength in the
1--5~$\mu$m region at a resolution of $\sim$16~km~s$^{-1}$ (two pixels).
The $\sim$80$\arcsec$ long slit was oriented north-south for all
observations. Sky measurements were obtained with the telescope nodded
600$\arcsec$ to the west.

On the first night, we centered the slit on the southern CO peak at J2000
RA = 12:01:55.0, Dec = -18:53:01.5, detecting weak line emission at
Pa$\beta$ and Br$\gamma$, but no emission at Br$\alpha$. On the following
two nights we positioned the slit 3$\arcsec$ west of the CO peak, which
corresponds to the location of strongest mid-infrared emission as measured
by ISO. There we detected all three recombination lines at four spatially
distinct locations along the slit (see Fig.~\ref{fig-1}), which we call
Apertures I-IV (in the order of north to south). Details of the locations
are provided in Table~1.

\begin{deluxetable}{llcrr}
\tabletypesize{\scriptsize}
\tablecolumns{5}
\tablecaption{Parameters of Hydrogen Line-emitting Sightlines}
\tablewidth{0pt}
\tablehead{
\multicolumn{1}{c}{Label} &
\multicolumn{1}{c}{Declination Range\tablenotemark{a}} &
\multicolumn{1}{c}{Solid Angle} &
\multicolumn{2}{c}{A$_{V}$ (from CO)\tablenotemark{b}} \\
\multicolumn{2}{c}{} &
\multicolumn{1}{c}{arcsec$^{2}$} &
\multicolumn{1}{c}{Red} &
\multicolumn{1}{c}{~~Blue}
}
\startdata
Ap I & -18:52:07.0 -- 18.2 & 11.2 & 18.0 & \nodata \\
Ap II & -18:52:27.4 -- 36.2 & ~8.8 & 15.3 & \nodata \\
Ap III & -18:52:44.8 -- 55.2 & 10.4 & 15.9 & 29.0 \\
Ap IV & -18:53:01.2 -- 09.8 & ~8.6 & 23.9 & 22.0 
\enddata
\tablenotetext{a}{At R.A. = 12:01:54.8 (J2000).}
\tablenotetext{b}{Visual extinction estimated from CO 1-0 column
desnity (see text). We use 1530 km s$^{-1}$ to separate the ``Blue'' and
``Red'' velocity components.}
\end{deluxetable}

\begin{deluxetable}{lccccccc}
\tabletypesize{\scriptsize}
\tablecolumns{8}
\tablecaption{Infrared Line Emission}
\tablewidth{0pc}
\tablehead{
\multicolumn{1}{c}{Line} &
\multicolumn{1}{c}{A$_{\lambda}$/A$_{V}$} &
\multicolumn{1}{c}{Exposure} &
\multicolumn{4}{c}{Flux (10$^{-18}$ W
m$^{-2}$)\tablenotemark{a}} &
\multicolumn{1}{c}{Observing} \\
\multicolumn{1}{c}{} &
\multicolumn{1}{c}{} &
\multicolumn{1}{c}{(sec)} &
\multicolumn{1}{c}{Ap~I} &
\multicolumn{1}{c}{Ap~II} &
\multicolumn{1}{c}{Ap~III} &
\multicolumn{1}{c}{Ap~IV} &
\multicolumn{1}{c}{dates (2000)}
}
\startdata
Pa $\beta$ (1.282$\micron$) & 0.270 & 7200\tablenotemark{b} &
5.1$\pm$0.3 & 3.9$\pm$0.2 & 5.3$\pm$0.3 & 4.6$\pm$0.3 & 02/29, 03/01
\\
Br $\gamma$ (2.166$\micron$) & 0.115 & 4800 & 1.06$\pm$0.06 &
1.08$\pm$0.04 & 1.72$\pm$0.06 & 1.41$\pm$0.03 & 02/29, 03/01 \\
Br $\alpha$ (4.052$\micron$) & 0.042 & 2600 & 4.7$\pm$0.8 & 4.0$\pm$0.5 &
10.0$\pm$1.4 & 5.8$\pm$0.8 & 03/01 
\enddata
\tablenotetext{a}{integrated over LSR velocities 1520--1660,
1550--1670, 1320--1495, and 1410--1580 km~s$^{-1}$, for Ap I, II, III,
and IV, respectively} \\
\tablenotetext{b}{Except 6600 for Ap~II}.
\end{deluxetable}

The data were wavelength-calibrated to an accuracy of 5~km~s$^{-1}$ using
spectra of arc lamps and night sky emission lines and flux-calibrated
using the A5V star HR~4405 (V=4.10), assumed to have the visible--infrared
colors given in \citet{tok99} for such stars. From the point-spread
functions along the slit we estimate that at the wavelength of each
recombination line 0.60~$\pm$~0.05 of the stellar flux entered the slit on
each night. Spectra were summed over $\sim$10 adjacent rows at each of the
four spatially distinct line-emitting regions. Data from the final two
nights were combined to yield the spectra shown in Fig.~2. The
figure also includes CO spectra obtained by L01 at those
locations. The basic results are summarized in Table~2.

\section{RESULTS}

\subsection{Spatial and Velocity Distributions}

Figure 2 shows broad recombination line emission, covering
100--200~km~s$^{-1}$, at each of the four locations along the slit. A
faint continuum is detected in some spectra. Ap~I and Ap~II are located
near local maxima of CO emission in the northern part of the interacting
region, where \citet{wilson00} and L01 have found largely redshifted CO
emission.
Ap~III and Ap~IV fall into the southern part of the interacting region
where both groups found overlapping red and blue components separated by
~$\sim$100~km~s$^{-1}$. 

\figcaption[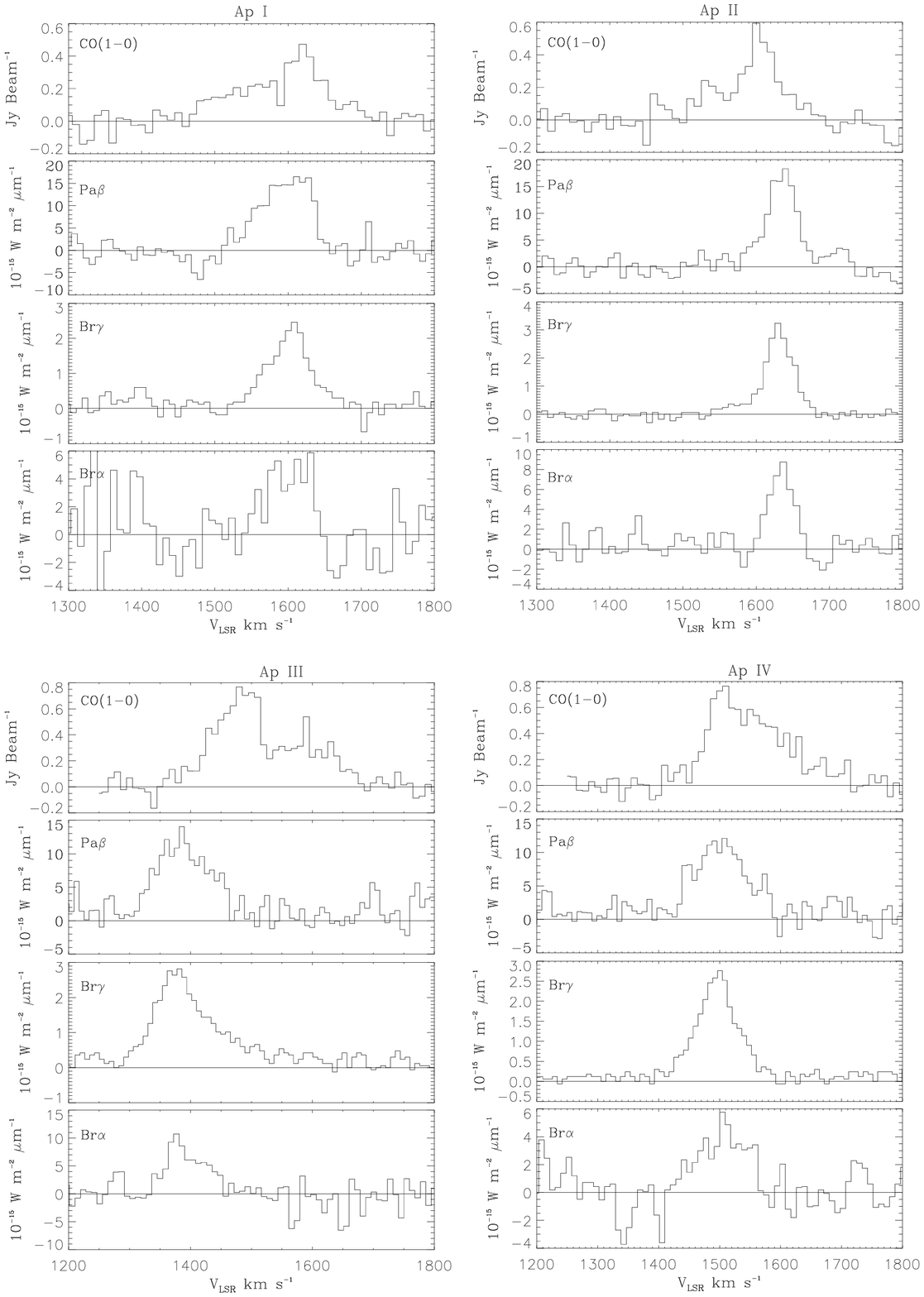]{Spectra at Apertures I-IV. The unit of flux
density for the infrared lines is 10$^{-15}$ Watts m$^{-2}$
$\mu$m$^{-1}$.\label{fig-2}}

In the two northern apertures the velocities and profiles of the infrared
lines generally match those of the CO emission. At Ap~I both the atomic
and molecular line profiles show peaks near V$_{LSR}$~=~1600~km~s$^{-1}$,
close to the velocities of the nuclei (the NASA/IPAC Extragalactic
Database gives radial velocities near 1640~km~s$^{-1}$ for both NGC~4038
and NGC~4039), and a blue asymmetry in the wings, and at Ap~II both show a
more symmetric line profile peaked near the same velocity. In the south,
however, there are distinct differences between the millimeter and
infrared velocity profiles. Most of the infrared line emission at Ap~III
occurs near V$_{LSR}$~=~1400~km~s$^{-1}$, $\sim$100~km~s$^{-1}$ to the
blue of the velocity of CO peak emission and in fact further blueshifted
than almost all of the CO. No infrared line emission is present near
1600~km~s$^{-1}$, the peak of the CO redshifted component. At Ap~IV the
velocities of peak emission are $\sim$1500~km~s$^{-1}$ for both the
infrared and millimeter lines, but little or no redshifted emission, which
is seen in CO near V$_{LSR}$~=~1600~km~s$^{-1}$, is detected in any of the
infrared lines.

At none of the locations are there more than marginally significant
differences between the profiles of the three infrared lines.  However,
the signal-to-noise ratios are not high, especially for Br$\alpha$, so
modest differences in profiles cannot be ruled out. In contrast, the
differences between the CO and hydrogen profiles are large, especially at
the southern positions.

\begin{deluxetable}{lrrrr}
\tablecolumns{5}
\tabletypesize{\scriptsize}
\tablewidth{0pt}
\tablecaption{Derived Extinctions}
\tablehead{
\multicolumn{1}{c}{Line Ratio} &
\multicolumn{4}{c}{A$_{V}$ (mag)} \\
 & Ap~I & Ap~II & Ap~III & Ap~IV
}
\startdata
Pa$\beta$/Br$\gamma$ & 1.4$\pm$0.5 &  3.5$\pm$0.5 & 4.5$\pm$0.5 &
4.1$\pm$0.5 \\
Pa$\beta$/Br$\alpha$ & 3.1$\pm$0.8 &  3.6$\pm$0.6 & 6.5$\pm$0.7 &
4.6$\pm$0.7 \\
Br$\gamma$/Br$\alpha$ & 6.7$\pm$2.5 & 4.0$\pm$1.6 & 10.7$\pm$1.9 &
5.8$\pm$1.6
\enddata
\end{deluxetable}

\subsection{Extinction}

The CO measurements of L01 and the infrared line measurements
reported here each provide a means for estimating extinction. Using the
mean CO to H$_{2}$ conversion factor determined for our Galaxy (e.g.,
Rand, Lord, \& Higdon 1999; Bryant \& Scoville 1999),
\begin{eqnarray*}
 \mbox{X} \equiv \frac{\mbox{N(H$_{2}$)}}{\mbox{I$_{\mbox{CO}}$}}
= 2.5\times 10^{20}  \mbox{H$_{2}$} ~ \mbox{cm$^{-2}$} 
\mbox{(K~km~s$^{-1}$)$^{-1}$} \\
= 6.2 \times 10^{11} \mbox{H$_{2}$} ~ \mbox{cm$^{-2}$}
\mbox{(Jy~km~s$^{-1}$)$^{-1}$},
\end{eqnarray*}
\noindent and the gas to dust ratio of \citet{BSD78},
\begin{eqnarray*}
 \mbox{N(H$_{2}$)} & = & 0.94\times 10^{21} \mbox{A$_{V}$} ~
\mbox{cm$^{-2}$~mag}, \end{eqnarray*} \noindent we determine that the
visual extinctions {\it through the molecular gas} at Ap~I and Ap~II are
18~mag and 15~mag, respectively. Those at Ap~III and Ap~IV are much
larger, A$_{V}$~$\sim$~45~mag and A$_{V}$~$\sim$~46~mag, respectively. All
of these estimates incorporate both interferometric and single dish
observations. It is known that in some starburst galaxies the conversion
factor is up to three times smaller than in our Galaxy
\citep{smi91,dow98}, and therefore it is possible that the extinctions are
proportionately less in the Antennae overlap region.

To derive the extinctions to the infrared line-emitting regions, we
assumed typical \ion{H}{2} region conditions (T$_{e}=10^{4}$ K,
n$_{e}=10^{4}$ cm$^{-3}$), for which the intrinsic ratios are 0.162,
0.0275, and 0.0777 for Pa$\beta$/H$\beta$, Br$\gamma$/H$\beta$, and
Br$\alpha$/H$\beta$ respectively \citep{hum87}. We employed the formula
obtained by \citet{rl85} for 0.9--3.4~$\micron$, along with an
interpolation of their $L$ and $M$ band extinctions, obtaining factors of
0.270 at Pa$\beta$, 0.115 at Br$\gamma$, and 0.042 at Br$\alpha$ relative
to visual extinction. We used the total integrated line fluxes as
described in Table~2; the overall general agreement of
the three infrared line profiles at each aperture suggests that at each
location the {\it detected} lines are attenuated by the same clouds.

The results are tabulated in Table~3. Significantly
different values are obtained for different pairs of lines at two of the
four locations. At all four locations the smallest value corresponds to
the shortest wavelength pair (Pa$\beta$/Br$\gamma$) and the largest value
corresponds to the longest wavelength pair (Br$\gamma$/Br$\alpha$). This
indicates that, rather than a single source of line emission and an
obscuring screen, there are distributions of line emission and attenuating
clouds along the lines of sight at each location \citep{pux91}.

\section{LOCATIONS OF THE CLOUDS}

The extinctions derived from the hydrogen recombination lines
(Table~3) are low compared to those derived from the CO
measurements (Table~1). This is the case even if the
extinctions through the molecular clouds are a factor of three less (as
discussed in section 3.2) except for the largest value derived for Ap~I.
The results indicate that most of the the observed recombination line flux
in each of the four apertures is emitted near the front surface of the
associated cloud complex and, moreover, for Ap~III and Ap~IV that complex
must be the only one that is absorbing and hence the one that is closer to
us. Although the velocities and profiles of the infrared lines at Ap~III
and Ap~IV do not closely match those of either molecular cloud, it is much
more likely that these blueshifted lines are associated with the
blueshifted molecular cloud than the redshifted one. Thus we conclude that
blueshifted molecular cloud is in the foreground.

A second approach makes use of the lack of redshifted line emission in
the region where the clouds overlap, although redshifted emission is
present in the northern region where the blueshifted cloud is absent. The
non-detection at Ap~III implies either that no redshifted recombination
line emission occurs there or that high extinction makes the infrared
lines unobservable. For the latter case, if we assume the same ratios of
infrared line emission to CO line emission in Ap~III as in Ap~I or Ap~II
we obtain rough lower limits for A$_{V}$ to the redshifted cloud. These
may be compared to the values of A$_{V}$ obtained from the CO line
strength (Table~1) to infer the geometry. We obtain
lower limits to A$_{V}$ at redshifted velocities of $\sim$8, $\sim$20,
and $\sim$27~mag for Pa$\beta$, Br$\gamma$, and Br$\alpha$, respectively.
The most restrictive of these, the Br$\gamma$ and Br$\alpha$ limits, rule
out extinction by the redshifted cloud {\it alone}, even if that emission
originates entirely behind it.  We conclude that both clouds would be
required to produce the extinction and, therefore, at Ap~III we again
infer that the blueshifted cloud complex is in front of the redshifted
complex. The above argument could be incorrect if the recombination line
emission from the red-component in Ap~III is intrinsically weak. At Ap~IV
it is more difficult than at Ap~III to clearly separate the two cloud
complexes based on the CO velocity profile, so this test is more
problematic than for Ap~III, although it tends to suggest the same
conclusion.

The significant blueshift of the infrared lines at Ap~III relative to the
blueshifted molecular component is apparent even at the highest resolution
($\sim$2$\arcsec$) CO measurements of L01. This nearly complete
lack of overlap in velocities is somewhat surprising. We propose two
possible explanations. 1. The ionized gas lies close to the front surface
of the blueshifted molecular cloud complex and is largely accelerated
outward and toward the observer. 2. The stars with which the ionized gas
is associated have a different velocity distribution than the molecular
cloud.

Several possibilities exist to test the correctness of our conclusions
about the relative positions of the molecular cloud complexes along the
line of sight. Measurements of the CO lines at higher angular
resolution would better match the spatial resolution of the infrared
measurements. Alternatively, spectral mapping of the infrared lines would
allow a better comparison with the present CO observations. Searches for
redshifted radio recombination line emission could determine if any
ionized gas, undetected at infrared wavelengths, is associated with the
redshifted molecular gas.

\section{SUMMARY AND CONCLUDING DISCUSSION}

We have obtained velocity-resolved spectra of three infrared hydrogen
recombination lines, covering a wide range of wavelengths, in the
interaction region of the Antennae galaxies, where millimeter CO
observations have revealed two distinct clouds of gas with widely
different radial velocities.  Using an empirical extinction formula,
standard recombination line ratios, and the standard gas-to-dust
conversion factor, we infer that the two molecular clouds currently are
moving away from each other. This is contrary to the conventional
expectation of cloud collisions leading to starburst activities (e.g.
\citet{wilson00}).

The age of the starburst in the interacting region has been estimated to
be 3--4 Myrs \citep{Hvan86,neff00}. If this starburst were due to cloud
collisions, then at present the clouds can only be separated (along the
line of sight) by 100 km s$^{-1}$ times a few Myrs, or $\sim$300~pc. The
cloud complexes are about one kpc in extent in the plane of the sky. If
they are roughly spherical they would still be physically overlapped in
the radial direction. Yet we observe two well-defined components moving
apart at 100 km s$^{-1}$. How this physical arrangement of the
molecular clouds might have occurred is not clear. There is no easy way to
determine the radial separation, if any, of the two molecular cloud
complexes, although in principle, one could model the interaction of the
gas in Arp 244 in sufficient detail to see if this physical situation is
plausible.  The morphology of the molecular cloud complexes
(Fig.~\ref{fig-1}) does not suggest physical proximity.

If the starbursts currently observed are not due to direct cloud
collisions, the physical mechanisms initiating them require further
investigation. Current observational evidence of molecular gas in
starburst regions points to two characteristic parameters: high column
density \citep{dow98,bry99} and high density \citep{solomon92,dow98}.  If
high column density and high density (both of which are naturally expected
in the process of gravitational collapse) are causal factors for
starbursts, then the issue reduces to identifying the mechanisms during
galaxy interaction that lead to the concentration of a large amount of
molecular gas in regions such as the interacting region in Arp 244.

\acknowledgements

We thank the staff of UKIRT for technical support. UKIRT is operated by
the Joint Astronomy Centre on behalf of the U.K. Particle Physics and
Astronomy Research Council.  This work is supported by the National
Science Council and the Academia Sinica in Taipei.  We thank the staff of
ASIAA for helpful discussions and suggestions.  This work also
constitutes part of the Master's degree thesis of M. C. Liang at the
National Tsing Hua University.

\end{document}